%% file: main.tex
\documentclass[10pt,twocolumn,twoside]{IEEEtran}

\usepackage{bm}

\usepackage[short]{optional}

\usepackage{cite}
   \usepackage[dvips]{graphicx}

\usepackage[cmex10]{amsmath}
\usepackage{algorithmicx}
\usepackage[ruled]{algorithm}
\usepackage{array}
\usepackage[tight,footnotesize]{subfigure}

\usepackage{graphicx, amssymb,subfigure,float}
\usepackage{times}
\usepackage[section]{placeins}
\usepackage{caption}
\usepackage{color}

\newtheorem{theorem}{Theorem}
\newtheorem{lemma}{Lemma}
\newtheorem{definition}{Definition}
\newtheorem{corollary}{Corollary}
\newtheorem{proposition}{Proposition}
\def\squareforqed{\hbox{\rlap{$\sqcap$}$\sqcup$}}
\def\qed{\ifmmode\squareforqed\else{\unskip\nobreak\hfil
\penalty50\hskip1em\null\nobreak\hfil\squareforqed
\parfillskip=0pt\finalhyphendemerits=0\endgraf}\fi}

\newcommand{\ignore}[1]{}

\newif\iftp \tpfalse

\begin{document}

\input{abstract}

\input{introduction}

\input{related}

\input{model}

\input{games}
\input{static}
\input{dynamic}
\input{simulation}

\input{conclusion}

\section*{Acknowledgements}
The effort described in this article was partially sponsored by the U.S. Army Research Laboratory Cyber Security Collaborative Research Alliance under Contract Number W911NF-13-2-0045.  The views and conclusions contained in this document are those of the authors, and should not be interpreted as representing the official policies, either expressed or implied, of the Army Research Laboratory or the U.S. Government. The U.S. Government is authorized to reproduce and distribute reprints for Government purposes, notwithstanding any copyright notation hereon.

\vspace{-1ex}
\bibliographystyle{ieeetr}
\bibliography{trust_SNSs_game}
\end{document}

%% file: abstract.tex
\title{Trust Exploitation and Attention Competition: A Game Theoretical Model}



\author{
    \IEEEauthorblockN{Hao Fu, Hongxing Li, Zizhan Zheng, Pengfei Hu, Prasant Mohapatra} \\
    \IEEEauthorblockA{Department of Computer Science, University of California, Davis, CA, USA.
    \\\texttt{\{haofu, honli, cszheng, pfhu, pmohapatra\}@ucdavis.edu}}
}

\maketitle
\begin{abstract}
The proliferation of Social Network Sites (SNSs) has greatly reformed the way of information dissemination, but also provided a new venue for hosts with impure motivations to disseminate malicious information. \emph{Social trust} is the basis for information dissemination in SNSs. Malicious nodes judiciously and dynamically make the balance between maintaining its social trust and selfishly maximizing its malicious gain over a long time-span. Studying the optimal response strategies for each malicious node could assist to design the best system maneuver so as to achieve the targeted level of overall malicious activities. In this paper, we propose an interaction-based social trust model, and formulate the maximization of long-term malicious gains of multiple competing nodes as a non-cooperative differential game. Through rigorous analysis, optimal response strategies are identified and the best system maneuver mechanism is presented. Extensive numerical studies further verify the analytical results.
\end{abstract}

%% file: introduction.tex
\section{Introduction}
\label{sec:introduction}
We have witnessed the prevailing useage of Social Network Sites (SNSs), including Facebook, Twitter and Google+, for the information sharing among users on their personal pages and the interaction with friends or followers\cite{ellison2007social}. While SNSs provide excellent platforms for information dissemination among millions of users \cite{boshmaf2011socialbot}, they also attract hosts with impure motivations to exploit their massive influence for malicious activities, such as spam, click fraud, identity theft and phishing \cite{thomas2011design}.

Unique feature of SNSs is that the information dissemination is primarily dependent on the \emph{social trust} among users, \cite{sherchan2013survey}, \emph{e.g.}, a user's post is more likely to be reposted by his/her followers instead of others with no social tie. As a result, the one-time gain from a malicious action is positively related with the social trust of the malicious user, \emph{i.e.}, the higher the social trust is, the more users will be influenced by the malicious action.

The social trust of a user reflects the confidence that this user will behave in an expected way, and can be evaluated by his/her frequency of non-malicious interactions with other users \cite{sherchan2013survey}. A positive interaction, \emph{e.g.}, posing a trustworthy news, will improve the social trust of the user leading to larger influence for information dissemination, while a negative interaction, \emph{e.g.}, maliciously spreading a rumor, will result in a degradation in the trust and hurting his/her potential of information dissemination in the future. Hence, for a malicious user aiming to maximize his/her overall personal benefits over a long time span, a tradeoff should be made between dynamically conducting positive and negative interactions with others, \emph{e.g.}, obtaining malicious gain through negative interactions while accumulating better trust by positive interactions for larger malicious gain later.

It is desirable to understand the malicious host's best action strategy towards this tradeoff, and to accordingly propose optimal system maneuver mechanism for the social trust management so as to confine the malicious activities in the system. However, it is non-trivial to find the optimal balance between positive and negative interactions so as to maximize the long term malicious gain: how can we quantify the impact of an action on the future malicious gains and judiciously conduct positive/negative actions dynamically?

The difficulty further escalates when we practically extend the problem of optimizing the malicious gain at one individual user to the picture of interplays among multiple malicious users, who compete for the social trust, \emph{i.e.}, interaction densities with other normal users, in order to selfishly maximize their own influence in information dissemination and thus malicious gains. Each action taken by an individual user will have an impact on the potential gain of other malicious users and vice versa. The following questions should be answered: how to evaluate the impact of an action on one's own and others' malicious gains in the future; what is the best strategy for each malicious user to dynamically adjust his/her positive/negative interactions in this competition?

Each user can be viewed as a node in the online social network. Our objective is to study the optimal response strategies of the malicious nodes in both single-node case and multiple-node case, respectively, such that we could find a better system maneuver accordingly in order to manage the trust evaluation and control the malicious activities. We propose an interaction-based \emph{social trust} evaluation model, and formulate the single-node case as an optimal control problem and the competition among multiple nodes as a non-cooperative differential game. Through rigorous analysis, we solve the optimal response strategies for each node in both cases, on the basis of which we identify the best system maneuver mechanism given any targeted level of overall malicious activities.

The main contributions of this paper can be summarized as follows.
\begin{itemize}
\item We investigate  \emph{social trust} and its impact on the malicious information dissemination in SNSs.
\item We propose a general framework to model the \emph{social trust} using the frequency of interactions in the SNSs. Based on this model, we gain the insight for the administrators of SNSs to control the overall malicious activity.
\item Through rigorous analysis, we identify the best response strategies for each node in both the single-node optimal control problem and the multiple-node differential game. Best system maneuver strategies are presented for each case, so as to maintain the overall malicious activities at any given level.
\item Extensive numerical studies further verify our analytical results in differential system settings.
\end{itemize}

The rest of the paper is organized as follows. We highlight the related work in Section~\ref{sec:related_work}.
In Section~\ref{sec:games}, we present our model on the social trust and the threats from malicious nodes. We solve the multiple-node differential game for both static and dynamic cases in Section~\ref{sec:games}. Numerical studies under different system settings are presented in Section~\ref{sec:simulation}. Finally, Section \ref{sec:conclusion} concludes the paper.

%% file: related.tex
\section{Related Work}
\label{sec:related_work}

\subsection{Social Trust and Trust Management}
As an emerging research topic, social trust in social networks has been extensively discussed in \cite{sherchan2013survey} and references therein. \ignore{Social trust is an assessment of the given host's honest intention in a social network. In order to create and maintain trust communities, many interaction-based trust models have been proposed to evaluate the trustworthiness of a given host in online social networks \cite{adali2010measuring}\cite{trifunovic2010social}.
Adali \emph{et al.}~\cite{adali2010measuring} propose a trust evaluation method based on communication behaviors, which can be further categorized into the popularity and the engagement trust.
Tricfunovic \emph{et al.}~\cite{trifunovic2010social} separate social trust into two classes, where implicit trust is a measure of frequency and duration of the contact among a pair of members.}
The application of trust frameworks and systems in social networks involves defending malicious activities, especially the spamming \cite{stringhini2010detecting}\cite{yang2011free}\cite{wang2014provenance}.
As mentioned in \cite{stringhini2010detecting}, the behaviors of spammers are getting stealthy to evade from existing detection techniques.
Yang \emph{et al.}~\cite{yang2011free} state that malicious hosts can dilute their vicious posts and raise the opportunities to survive through mixing normal content with malicious content.
To effectively eliminate the threat from spammers, 
Wang \emph{et al.}~\cite{wang2014provenance} design an trust based collaborative spam mitigation system.

\subsection{Game Theory in Cybersecurity}
As a technique that naturally supports modeling decision-making for multiple agents, game theory has been extensively applied in security area.
Hu~\emph{et al.}~\cite{hu2015dynamic} and Feng~\emph{et al.}~\cite{xiaotao2016, xiaotaomilcom} propose dynamic game models to analyze the interplay among attacker, defender and insider.
Omic~\emph{et al.}~\cite{omic2009protecting} combine the epidemic model with game theory in order to derive the optimal protection mechanism against infection,
whereas Zhu~\emph{et al.}~\cite{6426481} utilize differential games to analyze the infection process.

To the best of our knowledge, our work is the first in literature as the application of game theory for online social trust, and presents provably optimal system maneuver mechanism for SNSs.

%% file: model.tex
\section{Social Trust and Threat Model}
\label{sec:threatModel}

In this section, we first define the \emph{social trust} in social networks and its dynamics based on the mutual \emph{positive interactions}. Next, we formulate the threats from malicious nodes, and discuss the problem models for optimal tradeoff between positive and negative interactions. Important notations are summarized in Table \ref{table: notation}.

\begin{center}
\begin{table}
  \centering
  \caption{Important notations.}\label{table: notation}
  \begin{tabular}{|c|p{6cm}|}
    \hline
    $\alpha_i(t)$ & rate of posting trustable information from $i$\\ \hline
    $\beta_i(t)$ & rate of posting malicious information from $i$\\ \hline
    $x_i(t)$ & fraction of online users who are interacting with $i$ at $t$\\ \hline
    $\dot{x_i}(t)$ & the evolving rate of $x_i$ at each time point\\ \hline
    $x_{i0}$ & initial value of the $x_i$\\ \hline
    $P_{i}(\cdot)$ & long-term profit gain of $i$ from negative activities\\ \hline
    $C_{i1}(\cdot)$ & long-term cost for positive activities of $i$\\ \hline
    $C_{i2}(\cdot)$ & long-term cost for negative activities of $i$\\ \hline
    $\alpha_{-i}(t)$ & action profile of positive activities for all players except $i$\\ \hline
    $\beta_{-i}(t)$ & action profile of negative activities for all players except $i$\\ \hline
  \end{tabular}
\vspace*{-4.5mm}
\end{table}
\end{center}\vspace{-4mm}

\subsection{Social trust}\label{subsec:social_trust}
As a measurement of the confidence that an entity will behave in an expected way, trust moves to the center of data dissemination in SNSs. To build a trust community where users provide healthy information and feel free to share with each other, an effective and convenient trust system is required.

The interactions between a pair of users provides a natural way to assess one's \textit{social trust} \cite{sherchan2013survey}. Users with high social trust draw more attention from others and involve high frequency of positive interaction with their neighbors, whereas un-trusted nodes get little attention and have limited influence of data dissemination over the SNSs. Current SNSs offer features to reflect one's social trust level based on users' reactions on the posted information, \emph{e.g.}, Facebook users normally click ``like'' or ``share'' if they are in a comfortable interaction and they could choose to report a spam if they feel offended by the content.

In this paper, we use a general model to characterize one's social trust to the rest of the social network. Let $N$ denote the total number of users in the social network. 
Let $X_i(t)$ denote the number of users that trust node $i$ at time $t$, which is a random variable in general. We model the social trust of node $i$ at as $x_i(t) = \mathbb{E}(X_i(t)/N)$, which evolves over time. Its dynamics is determined by its initial value $x_{i0}\in [0,1]$, which is a constant, and its actions on disseminating trustable/malicous information as discussed below. Alternatively, we can consider $x_i(t)$ as the fraction of nodes that interact with node $i$ (assuming a node only interacts with the set of nodes that it trusts).

\subsection{Dynamics of Social Trust}
A malicious node delivers malicious content to as many users as possible for a profit. However, it does not target at a one-time profit from disseminating the malicious information. Instead, it tries to persistently make profits over a long time-span by continuously spreading malicious information. As discussed previously, social trust determines the influence of the information on the SNSs. Hence, a malicious node does not consistently provide pure baleful content to avoid diminishing its social trust and its information influence for later malicious actions. Instead, it moves stealthily by mixing good content with malicious content. It can either mix both type of content into one post or by posting these two in separate claims\cite{stringhini2010detecting}\cite{yang2011free}.
By doing so, it maintains an acceptable level of social trust, and makes a balance between its instantaneous malicious gain of current action and its future profits.

\vspace{1ex}
\noindent{\bf Single malicious node:} Let us first consider the case of a single malicious node. Consider a malicious node $i$ that posts some content $c(t)$ at time $t$.
We model the impact of the content on the dynamics of social trust of node $i$ by a pair of transition probabilities. Let $p_1(c(t),\delta)$ denote the probability that a node distrusting $i$ at time $t$ becomes trusting $i$ at time $t+\delta$ after the content is posted for a small time period $\delta$, which depends on both the content posted and $\delta$. Similarly, let $p_2(c(t),\delta)$ denote the probability that a node trusting $i$ at time $t$ becomes distrusting $i$ at time $t+\delta$. 
Intuitively, $p_1$ models the negative influence of malicious content through direct interaction with node $i$, and $p_2$ models the positive influence of benign content that propagates indirectly, e.g., through the ``word-of-mouth" effect. In both cases, the influence is assumed to be independent across nodes. It follows that
\begin{align}
\mathbb{E}(X_i(t+\delta) - X_i(t)|X_i(t)) = p_1(&c(t),\delta)(N-X_i(t)) \nonumber \\
&- p_2(c(t),\delta)X_i(t),
\end{align}
\noindent Taking the expectation (with respect to $X_i(t)$) of both sides, we have
\begin{align}
x_i(t+\delta) - x_i(t) = p_1(c(t),\delta)(1-x_i(t)) - p_2(c(t),\delta)x_i(t).
\end{align}

\noindent Dividing both sides by $\delta$ and letting $\delta \rightarrow 0$, we obtain the following dynamics of social trust of node $i$,
\begin{align}\label{eq: odx}
   \dot{x_i}(t) & = \frac{dx_i(t)}{dt}
	 = \alpha_i(t) (1 - x_i(t))
    -  \beta_i(t)x_i(t),\\
   x_i(0) & = x_{i0},\notag
\end{align}
where $\alpha_i(t) = \lim_{\delta \rightarrow 0} \frac{p_1(c(t),\delta)}{\delta}$ and $\beta_i(t) = \lim_{\delta \rightarrow 0} \frac{p_2(c(t),\delta)}{\delta}$, which are assumed to exist.
Instead of modeling the details of $p_1$ and $p_2$, we consider $(\alpha_i(t), \beta_i(t))$ as the strategy of node $i$ in this work. We note that the differential equation is intuitive by itself. In particular, $\alpha_i(t) (1 - x_i(t))$ can be viewed as the social trust gained by posting trustable information that has positive response from $1-x_i(t)$ (the share that is originally not positively interacting with node $i$); while $\beta_i(t)x_i(t)$ reflects the loss of social trust because of disseminating malicious content to $x_i(t)$ (the share that is originally positively interacting with node $i$). To simplify the description, we normalize $\alpha_i(t)$ and $\beta_i(t)$ so that  $\alpha_i(t) + \beta_i(t) =1$.

\vspace{1ex}
\noindent{\bf Multiple malicious nodes:} Next, we consider the coexistence of multiple malicious nodes in the SNSs and the competition among them. 
As mentioned previously, social trust can be viewed as the frequency of interactions among users. In a continuous-time environment as in real-life applications, a content viewer in SNSs only involves in an effective interaction with one content provider at one time point. For instance, a user cannot click ``like'' for two separate posts concurrently at exactly the same time.
Moreover, each online user has a limit \textit{budget of attention} as suggested in \cite{jiang2013optimally}.
The notion budget of attention quantifies the constraint on one's frequency of pulling content from the neighbors.
Since attention is the foundation and the necessary condition for interaction, we extend the concept so as to characterize the upper bound exists on user's interaction rate.

\begin{definition}[Budget of interaction]
Budget of interaction is a constrained rate of a user that quantifies all kinds of its positive actions, which exclusively happen in continuous time at a social network site.
\end{definition}

That is to say, malicious nodes have to compete with each other to gain social trust from their potential victims in order to maximize their individual profits.

Given $n$ competing malicious nodes ($n > 1$) in a online social network, we assume that the sum of their social trust should be upper-bounded by the total interactions in the entire network, which is normalized to 1. That is,
\begin{align}
\sum_i{x_i (t)} \leq 1.\label{eq:trust_bound}
\end{align}

Different from the previous single-node case, the dynamics of node $i$'s social trust should consider the joint actions of all the malicious nodes and be formulated as follows,
\begin{align}\label{eq: dx}
   \dot{x_{i}}(t)
	 = & \alpha_{i}(t) (1 - x_{i}(t)) - \sum_{j \in -i}\alpha_{j}(t)x_{i}(t) -  \beta_{i}(t)x_{i}(t), \notag \\ 
x_i(0) = & x_{i0} 
\end{align}
where $\alpha_{i}(t) (1 - x_{i}(t))$ and $\beta_{i}(t)x_{i}(t)$ have the same meaning as that in its counterpart with single malicious node; while $\sum_{j \in -i}{}\alpha_{j}(t)x_{i}(t)$ denotes the accumulated loss rate of social trust, that is obtained by other malicious nodes, \emph{i.e.}, $j\in -i$, who post trustable information and attract the share that is originally positively interacting with node $i$.
Above derivative equation captures the effect of "word-of-mouth" and diffusion progress, which are often used for advertising and marketing in economics field \cite{jorgensen1982survey}.

We can find from Eqn.~(\ref{eq:trust_bound}) and Eqn.~(\ref{eq: dx}) that, each malicious node has to compete with each other for higher social trust, which leads to higher profit gain accordingly to Eqn.~(\ref{eq:pay}).

\subsection{Payoff and Cost Functions for Malicious Nodes}\label{subsec:model_cost}
The instantaneous malicious profit of node $i$ at time $t$ should be proportional to its malicious activity rate $\beta_i(t)$ and its social trust $x_i(t)$, \emph{i.e.}, the amount of interactions could be influenced. Hence, the long-term profit gain $P_i$ for node $i$ is defined as follows,
\begin{align}
    &P_i = \lim_{T \to \infty} \frac{1}{T} \int_0^T p_i\beta_i(t) x_i(t) dt,
    \label{eq:pay}
\end{align}
where $p_i$ is the unit malicious profit for node $i$, with a positive constant value.

However, every activity comes with an operational cost. Both positive activity $\alpha_i(t)$ and negative activity $\beta_i(t)$ consume money in manpower at the malicious node.
As commonly applied in literature \cite{6426481}\cite{jorgensen1982survey} \cite{han2012game} , we utilize the quadratic cost function to capture the instantaneous operational costs. The long-term costs for positive activities, $C_{i1}$ and negative activities, $C_{i2}$, are evaluated as follows,
\begin{align}
    &C_{i1} =  \lim_{T \to \infty} \frac{1}{T} \int_0^T q_i\alpha_i^2(t)  dt,
    \label{eq:cost1}\\
    &C_{i2} = \lim_{T \to \infty} \frac{1}{T} \int_0^T r_i \beta_i^2(t) dt,
    \label{eq:cost2}
\end{align}
where $q_i$ and $r_i$ are the unit cost of providing trustable content and the unit penalty for each malicious activity, respectively.
$p_i$, $q_i$ and $r_i$ are all positive.

To sum up, the net profit for malicious node $i$ is
\begin{align}
    P_i - C_{i1} - C_{i2}.
    \label{eq:net_profit}
\end{align}

In the case of multiple malicious nodes, each of the malicious nodes acts independently and selfishly to maximize its individual net profit as defined in Eqn.~(\ref{eq:net_profit}).

\subsection{System Maneuver}\label{subsec:maneuver}
The objective of this paper is to find the optimal system maneuver mechanism, \emph{i.e.}, configuration of the system parameters, in order to control the overall malicious activity within the targeted level.

The overall malicious activity is defined as i) $\beta_i(t)$ for the single-node case; and ii) $\sum_{i\in [1, n]}\beta_i(t)$ for the multiple-node case, when $\beta_i(t)$ has converged to its optimal strategy.

As for the system administrator, it can adjust the value of $r_i$, which could be the unit penalty for malicious activities of node $i$, at the start of the system so as to achieve its targeted level of overall malicious activity. Note that $p_i$ and $q_i$ are constants that are only related with the malicious node's setting while not controllable by the system administrator.

%% file: optimal.tex
\section{The Single Malicious Host}
\label{sec:oc}
In this section, we first analyze the behavior of a single malicious host in SNSs, which is an optimal control problem as defined in Sec.~\ref{sec:threatModel}. We propose optimal solutions to two cases: the static case and the dynamic case, with details in Sec.~\ref{subsec:ostatic} and Sec.~\ref{subsec:odynamic}, respectively. Based on the optimal controls for the malicious host, we present the method to determine the system maneuver for controlling the malicious activity into any targeted level.


%% file: ostatic.tex
\subsection{Static Case}\label{subsec:ostatic}
In this subsection, we analyze the behavior of a single malicious host in the static case, where the rates of trustable and malicious actions are time-invariant and pre-configured before the deployment of the system.
The malicious host aims to maximize its own net profit with optimal control.
Based on the definition of its net profit as in Eqn.~(\ref{eq:net_profit}) and the dynamics of its social trust as in Eqn.~(\ref{eq: dx}), we can have the profit-maximization problem at each host $i$ as follows (for simplicity, we denote $\alpha_i(t)$ and $\beta_i(t)$ as $\alpha_i$ and $\beta_i$ since they are time-invariant in this subsection),
\begin{align}\label{eq: osp1}
	\text{max} &~~ \lim_{T \to \infty} \frac{1}{T} \int_0^T p_i\beta_i x_i(t) - q_i \alpha_i^2  - r_i \beta_i^2 dt\\
	\text{s.t.}& ~~ \dot{x}_i = \alpha_i (1 - x_i(t))
    -  \beta_i x_i(t), \quad x_i(0) = x_{i0}, \notag\\
    &~~\alpha_i,\beta_i \in [0,1], \quad \alpha_i+\beta_i=1. \notag
\end{align}


Since $\beta_i$ can be directly computed through $\alpha_i$, we use $\alpha_i$ to denote the action of the malicious host. The optimal control problem can be converted into the following form,
\begin{align}\label{eq: osp}
	\text{max} &~~ \lim_{T \to \infty} \frac{1}{T} \int_0^T p_i(1-\alpha_i)x_i(t)- r_i(1-\alpha_i)^2 - q_i \alpha_i^2 d t \\
	\text{s.t.}& ~~ \dot{x}_i = \alpha_i - x_i(t), \quad x_i(0) = x_{i0}, \label{eq: odx1}\\
    &~~\alpha_i \in [0,1] \notag
\end{align}

By solving ordinary differential equation (\ref{eq: odx1}), we can acquire the fraction of online users who involve positive interaction with the given malicious host in the SNS at time $t$ as
\begin{align}
	x_i(t)
	= & e^{-t}x_{i0}
	+ \alpha_i(1-e^{-t}).
\end{align}

Substituting $x_i(t)$ into Eqn.~(\ref{eq: osp}) and Eqn.~(\ref{eq: odx1}), the optimal control problem for the malicious host can be further simplified into
\begin{align}
	\text{max} & ~~p_i(1-\alpha_i)\alpha_i - r_i(1-\alpha_i)^2 - q_i \alpha_i^2,\\
	 s.t.& ~~\alpha_i \in [0,1]. \notag
\end{align}

We solve the above optimization problem with the following theorem.

\begin{theorem}\label{the: os}
    Consider the static case of single malicious host problem, an unique optimal control exists in [0,1] and is given by
	\begin{align*}
      \begin{cases}
		\alpha^* = \frac{p_i + 2r_i }{2(p_i + q_i + r_i)},\\	
        \beta^* = \frac{p_i + 2q_i}{2(p_i + q_i + r_i)},
        \end{cases}
	\end{align*}
\end{theorem}

\begin{IEEEproof}
	We define the objective function $F(\alpha_i) =  p(1-\alpha_i)\alpha_i - r_i(1-\alpha_i)^2 - q_i \alpha_i^2$.
	Taking the second order derivative of $F(\alpha)$ we have
    \begin{align*}
		F''(\alpha_i) =& -2p_i - 2(q_i + r_i).&
    \end{align*}
	As $p$, $q$ and $r$ can only be positive values,
    the second order derivative $F''(\alpha_i) < 0$ and it reflects the concavity of $F(\alpha)$.
    Then checking the value of first order derivative at the end points,
    \begin{align*}
    \begin{cases}
        &F'(0) = 2r_i + p_i > 0, \\
        &F'(1) = -2q_i - p_i < 0,
        \end{cases}
    \end{align*}
    which reflects that the uniqueness of optimal solution $\alpha_i^*$ and it exists in $(0, 1)$.
	The value of $\alpha_i^*$ can be directly computed from the first order derivative of $F(\alpha)$,
	\begin{align*}
		&F'(\alpha_i) = p_i + 2r_i -2(p_i + q_i + r_i)\alpha_i = 0.&
	\end{align*}
\end{IEEEproof}


With Theorem \ref{the: os}, we can have the optimal system maneuver with the following theorem.

\begin{theorem}\label{theorem:os_maneuver}
    Given a targeted level of malicious activity $\beta_i$, a unique optimal system maneuver exists for $r_i$ in the static single malicious host case and is given by
	\begin{align*}
			r_i^* = \frac{p_i + 2q_i}{2 \beta_i} - p_i - q_i.
	\end{align*}
\end{theorem}

The proof to this theorem can be trivially derived from Theorem \ref{the: os}, and omitted for limited space.

%% file: odynamic.tex
\subsection{Dynamic Case}\label{subsec:odynamic}
In this subsection, we analyze the dynamic case for the optimal control problem of single malicious host, where the control rates, \emph{i.e.}, $\alpha_i(t)$ and $\beta_i(t)$, are dynamically decided and adapted to the system status, \emph{i.e.}, $x_i(t)$, at time $t$.

The optimal control problem is defined the same as follows,
\begin{align}\label{eq: odp}
	\text{max} &~~ \lim_{T \to \infty} \frac{1}{T} \int_0^T p_i\beta_i x_i(t) - q_i \alpha_i^2(t)  - r_i \beta_i^2(t) dt\\
	\text{s.t.}& ~~ \dot{x}_i(t) = \alpha_i(t) (1 - x_i(t))
    -  \beta_i(t) x_i(t), \quad x_i(0) = x_{i0}, \notag\\
    &~~\alpha_i(t),\beta_i(t) \in [0,1], \quad \alpha_i(t)+\beta_i(t)=1. \notag
\end{align}

Let functions $f(\cdot)$ and $g(\cdot)$ denote the dynamics of social trust and the instantaneous net profit, respectively. We have that,
\begin{align*}
f(x_i(t), \alpha_i(t), \beta_i(t)) & = \dot{x}_i(t) = \alpha_i(t) (1 - x_i(t)) - \beta_i(t)x_i(t),\\
g(x_i(t), \alpha_i(t), \beta_i(t)) & =p_i\beta_i x_i(t) - q_i \alpha_i^2(t)  - r_i \beta_i^2(t).
\end{align*}


\begin{theorem}
Consider the dynamic single malicious host optimal control problem described in (\ref{eq: odp}), there exists an unique optimal control solution at $t$.
\end{theorem}
\begin{IEEEproof}
To obtain the optimal control, we utilize the Pontryagin Maximum Principle\cite{han2012game}.
We define the Hamiltonian function $H$ as
\begin{flalign}\label{eq: oH}
	&H(\alpha_i(t), \beta_i(t), x_i(t), \lambda_i(t)) = \notag\\
&\lambda_i(t)f(x_i(t), \alpha_i(t), \beta_i(t)) + g(x_i(t), \alpha_i(t), \beta_i(t))
\end{flalign}
where $\lambda_i(t)$ is the \textit{costate} (or \textit{adjoint}) variable attached to state variable $x_i(t)$.

Since $\beta_i(t)$ can be substituted by $1 - \alpha_i(t)$, we simply the problem as finding a control $\alpha^*_i(t)$ that satisfies
\begin{align*}
    \alpha_i^*(t) = \text{argmax} \  H(\alpha_i(t), x_i(t), \lambda_i(t)).
\end{align*}
The state space $[0, 1]$ is a compact set.
Also, the Hamiltonian function $H$ is concave and continuously differentiable with respect to $\alpha_i(t)$.
Thus, an unique optimal control exists.
\end{IEEEproof}

We move forward to solve for $\alpha_i^*(t)$ for each time instance.

\begin{lemma}
The optimal control solution for the dynamic single malicious host case is
\begin{align}\label{eq: oa*}
\alpha_i^*(t) =
\begin{cases}
\frac{\lambda_i(t) - p_ix_i(t) + 2r_i}{2(q_i + r_i)} & \lambda_i(t) > p_ix_i(t) - 2r_i,\\
0 & \text{otherwise.}
\end{cases}
\end{align}
\end{lemma}
\begin{IEEEproof}
From Eqn.~(\ref{eq: oH}), we can rewrite the $H$ as
\begin{align}
&H(\alpha_i(t), x_i(t), \lambda_i(t)) =  \lambda_i(t) (\alpha_i(t) - x_i(t)) \notag \\
			  &+ p_i(1 - \alpha_i(t))x_i(t) - q_i \alpha_i^2(t) - r_i (1 - \alpha_i(t))^2(t),
\end{align}
Solving the following first order condition,
\begin{align}\label{eq: ohs}
	\frac{\partial H_i}{\partial \alpha_i} = -2\alpha_i(t) (q_i + r_i) + \lambda_i(t) - p_i x_i(t) + 2r_i = 0,
\end{align}
we acquire the optimal solution as Eqn.~(\ref{eq: oa*}).
\end{IEEEproof}
At each time instance, we are able to solve for explicit value of $\alpha^*_i(t)$ through numerical approach from following Pontryagin necessary conditions
\begin{align}\label{eq: ohs}
	\frac{\partial H_i}{\partial \alpha_i} = 0,
\end{align}
\begin{align}\label{eq: oadj}
    - \frac{\partial H_i}{\partial x_i} = \dot{\lambda_i}.
\end{align}

When $x_i(t)$ reaches at its steady state, we can solve the costate variable analytically.


\begin{theorem}\label{theorem: od}
	As the state variable $x_i(t)$ of the dynamic single malicious host case converges to steady status, the optimal control is
	\begin{align*}
\begin{cases}
		\alpha_i^*(t) = \frac{p_i + 2r_i }{2(p_i + q_i + r_i)},\\	
        \beta_i^*(t) = \frac{p_i + 2q_i}{2(p_i + q_i + r_i)},
        \end{cases}
	\end{align*}
	which coincides with the solution for the static case of single malicious host optimal control problem.
\end{theorem}
\begin{IEEEproof}
Differentiating Eqn.~(\ref{eq: oa*}) and recalling the adjoint equation Eqn.~(\ref{eq: oadj}), we obtain
\begin{align*}
	\dot{\alpha_i}(t) &= \frac{1}{2(q_i + r_i)}(\dot{\lambda_i}(t)-p\dot{x_i}(t))& \notag \\
							&=\frac{1}{2(q_i + r_i)}(\lambda_i(t) - p_i x_i(t) + 2r_i - p_i(\alpha_i(t) - x_i(t))).&
\end{align*}

Then substitute $\lambda_i(t) = 2(q_i + r_i)\alpha_i^*(t) + p_ix_i(t) - 2r_i$ derived from Eqn.~(\ref{eq: oa*}), which yields
\begin{align*}
	\dot{\alpha_i}(t) 		 &=\frac{1}{2(q_i + r_i)}(2\alpha_i(t)(q_i + r_i)+2p_i x_i(t)-2r_i-p_i)&\\
						  &=0&
\end{align*}
in
\begin{align}\label{eq: oas}
	\alpha_i^*(t) = \frac{p_i + 2r_i - 2p_i x_i(t)}{2(q_i+r_i)}.
\end{align}
Since $x_i(t)$ finally converge to stationary status in open-loop solution, we have
\begin{align}{\label{eq: staCon}}
	\dot{x_i}(t) = 0.
\end{align}
From (\ref{eq: odx}), we then have
\begin{align}{\label{eq: staX}}
	x(t) = \alpha_i(t),
\end{align}
which can be plugged back to (\ref{eq: oas}) to solve for $\alpha^*(t)$.
\end{IEEEproof}

With Theorem \ref{theorem: od}, we can have the optimal system maneuver with the following theorem.

\begin{theorem}\label{theorem:od_maneuver}
    Given a targeted level of malicious activity $\beta_i$ when the system is in its stable status, a unique optimal system maneuver exists for $r_i$ in the dynamic single malicious node case and is given by
	\begin{align*}
			r_i^* = \frac{p_i + 2q_i}{2 \beta_i} - p_i - q_i.
	\end{align*}
\end{theorem}

The proof to this theorem can be trivially derived from Theorem \ref{theorem:od_maneuver}, and omitted for limited space.

\vspace{1mm}
\noindent \textbf{Remarks}: Since $p_i$ (the unit malicious profit for host $i$) and $q_i$ (the unit cost of providing trustable information for host $i$) are uncontrollable parameters for the system administrator, the value of $r_i^*$ should be set inversely proportional to the targeted malicious activity level $\beta_i$. Intuitively, the system administrator should increase the unit penalty for each malicious activity when it targets on a lower malicious activity. The contribution of Theorem \ref{theorem:od_maneuver} is deriving the analytical relation between the targeted malicious activity level and its corresponding system setting.

%% file: games.tex
\section{Social Trust Games}
\label{sec:games}
In this section, we study the competition among multiple malicious nodes and identify the best response strategy for each node. The competition is formulated into a non-cooperative differential game \cite{han2012game} that is continuously played among nodes. Note that the optimal control problem for the single malicious node setting can be easily derived from the game result and its result is given in our online technical report \cite{report}, since it can be viewed as a degenerate case of the differential game.

For each malicious node $i \in \{1, 2, ..., n\}$, it solves a profit-maximization problem in the game as follows,
\begin{align}\label{eq: ddsp1}
	\text{max} &~~ J_i(\alpha_i(t), \beta_i(t), \alpha_{-i}(t), \beta_{-i}(t)) \notag \\
&~~= \lim_{T \to \infty} \frac{1}{T} \int_0^T p_i\beta_i(t) x_i(t) - q_i \alpha_i^2(t)  - r_i \beta_i^2(t)dt\\
	\text{s.t.}& ~~ \dot{x}_i(t) = \alpha_{i}(t) (1 - x_{i}(t)) - \sum_{j \in -i}\alpha_{j}(t)x_{i}(t)  -  \beta_{i}(t)x_{i}(t), \notag \\
    &~~x_i(0) = x_{i0}, \ \alpha_i(t),\beta_i(t) \in [0,1], \ \alpha_i(t)+\beta_i(t)=1, \notag
\end{align}

We denote $\Phi(t) = \{\alpha_i(t), \alpha_{-i}(t); \beta_i(t), \beta_{-i}(t) \}$ as the strategy profile,
where $\{\alpha_{-i}(t), \beta_{-i}(t)\}$ is the action set of malicious nodes other than $i$.
$\phi_i(t) = \{\alpha_i(t), \beta_i(t)\}$ constitutes the strategy of $i$.
Our objective is to derive the {\it open-loop} Nash Equilibrium (NE) defined as follows.

\begin{definition}
	Consider the game described by Eqn.~(\ref{eq: ddsp}).
	The strategy profile $\Phi^*(t) = \{\phi_1^*(t) , ..., \phi_n^*(t) \}$ constitutes a Nash equilibrium solution if and only if,
    all following inequalities are satisfied
    \begin{align*}
        J_1(\phi_1^*(t), ..., \phi^*_n(t)) &\geq J_1(\phi_1(t), ..., \phi_n^*(t)),\\
		& \vdots \\
        J_n(\phi_i^*(t), ..., \phi^*_n(t)) &\geq J_n(\phi_1^*(t), ..., \phi_n^*(t)).
    \end{align*}
\end{definition}

Note that it is unrealistic for a malicious node to reveal its state to the competitors as the game evolves. Therefore, we consider the open-loop information structure in the game, which means that the players do not acquire further information except the common knowledge of the state vector at initial time $t = 0$ \cite{han2012game}.


%% file: static.tex
\subsection{Static Case}\label{subsec:static}
We first analyze the static scenario of multiple competing malicious nodes, where
the activity variables of all malicious nodes, \emph{i.e.}, $\alpha_i(t)$ and $\beta_i(t)$ remain unchanged during the runtime of the game. The goal of each malicious node is to maximize the individual net profit through choosing its optimal action before the game starts. Based on the definition of its net profit as in Eqn.~(\ref{eq:net_profit}) and the dynamics of its social trust as in Eqn.~(\ref{eq: odx}), we can have the optimal control problem as follows (for simplicity, we denote $\alpha_i(t)$ and $\beta_i(t)$ as $\alpha_i$ and $\beta_i$ since they are time-invariant in this subsection),
\begin{align}\label{eq: dsp}
	\text{max} &~~ J_i(\alpha_i, \alpha_{-i}) = \lim_{T \to \infty} \frac{1}{T} \int_0^T p_i(1-\alpha_i)x_i(t)- r_i(1-\alpha_i)^2\notag\\
&~~\hspace{24ex}- q_i \alpha_i^2 dt \\
	\text{s.t.}& ~~ \dot{x}_i = \alpha_{i} - x_{i}(t) - \sum_{j \in -i}\alpha_{j}x_{i}(t),~~ x_i(0) = x_{i0}, \label{eq: dx1}\\
    &~~\alpha_i \in [0,1] \notag
\end{align}
\noindent where we have used the fact that $\alpha_i+\beta_i=1$.

We obtain the fraction of users who involves positive interaction with the malicious node $i$ in the SNS at time $t$ through solving the ODE Eqn.~(\ref{eq: dx1}).
\begin{align}
	x_i(t)
	= & e^{-(1 + \sum_{j \in -i} \alpha_j)t}x_{i0} \\
	&+ \frac{\alpha_i}{1 + \sum_{j \in -i} \alpha_j} (1 - e^{-(1 + \sum_{j \in -i} \alpha_j)t}) \notag
\end{align}

Substituting the above $x_i(t)$ into the profit-maximization problem as defined in Eqn.~(\ref{eq: dsp}) and Eqn.~(\ref{eq: dx1}) , we can simplify the problem into,
\begin{align}\label{eq: Fs}
	\text{max}  &~~\frac{p_i(1-\alpha_i)\alpha_i}{1+\sum_{j \in -i}\alpha_{j}} - r_i(1-\alpha_i)^2
	- q_i \alpha_i^2\\
	s.t.& ~~\quad \alpha_i \in [0,1] & \notag
\end{align}

\iftp
\else
We can derive the best response of the malicious node $i$ by analyzing the structure of Eqn.~(\ref{eq: Fs}) and the proof is given in our online technical report~\cite{report}.
\fi
\begin{proposition}\label{the: s}
    For the static case of multiple competing malicious nodes, the best response for the malicious node $i$, where $i = 1, ..., n$ is given by
	\begin{align}\label{equ: a*}
            \alpha^*_i = \frac{p_i + 2r_i(1 + \sum_{j \in -i}\alpha_j)}
			{2[p_i+q_i+r_i + (q_i + r_i)\sum_{j \in -i}\alpha_j]}
	\end{align}
\end{proposition}

\begin{theorem}
There exists a Nash equilibrium for the static social trust game.
\end{theorem}
\begin{IEEEproof}
Let $B_i(a_{-i}) = a_i^*: [0,1] \rightarrow [0,1]$ be the best response function of $i$.
The action set [0, 1] is compact and convex.
Also, the best response function $B_i$ is continuous over [0, 1].
Thus, there exists a fixed point that satisfies the equation $\alpha_i^* = B_i(\alpha_i^*)$ based on Brouwer's fixed point theorem\cite{border1990fixed}.
Since a NE satisfies the fixed point equation, we prove the existence of NE.
\end{IEEEproof}

%% file: dynamic.tex
\subsection{Dynamic Case}\label{subsec:dynamic}
In this subsection, we analyze the dynamic case for the game among multiple malicious nodes.
The general term of dynamics has been described in Eqn.~(\ref{eq: dx}).
For each malicious node $i \in \{1, 2, ..., n\}$, it solves a profit-maximization problem in the game as follows (using the fact that $\alpha_i(t)+\beta_i(t)=1$),
\begin{align}\label{eq: ddsp}
	\text{max} &~~ J_i(\alpha_i, \alpha_{-i}) = \lim_{T \to \infty} \frac{1}{T} \int_0^T p_i(1-\alpha_i(t)) x_i(t) - q_i \alpha_i^2(t)\notag\\
  &~~\hspace{24ex}-r_i (1-\alpha_i(t))^2(t) dt\\
	\text{s.t.}& ~~ \dot{x}_i(t) = \alpha_{i}(t) - x_{i}(t) - \sum_{j \in -i}\alpha_{j}(t)x_{i}(t), \notag \\
    &~~x_i(0) = x_{i0}, \quad \alpha_i(t)\in [0,1]. \notag
\end{align}
We follow the procedure in \cite{cellini2007differential} to look for the open-loop Nash equilibrium.
\begin{lemma}\label{the: d}
    For the dynamic case of multiple competing malicious nodes, the best response of the malicious node $i = 1, ..., n)$ is given by
    \begin{align}{\label{eq: a*}}
	\alpha_i^*(t) =
\begin{cases}
\frac{\lambda_i(t) - p_ix_i(t) + 2r_i}{2(q_i + r_i)}& \lambda_i(t) > p_i x(t) - 2r_i,\\
0 & \text{otherwise.}
\end{cases}
\end{align}
\end{lemma}
\begin{IEEEproof}
To obtain the best response of each malicious nodes, we solve an optimal control problem for $i$.
The Hamiltonian function of $i$ is denoted by $H_i$ as:
\begin{align}
    H_i(\alpha_i, &\alpha_{-i}, t) =  \lambda_i(t) (\alpha_{i}(t) - x_{i}(t) - \sum_{j \in -i}\alpha_{j}(t)x_{i}(t))\notag\\
	&+ p_i(1 - \alpha_i(t))x_i(t) - q_i \alpha_i^2(t) - r_i (1 - \alpha_i)^2(t) \notag
\end{align}
where $\lambda_i(t)$ is the co-state variable attached to $x_i(t)$.

We apply the Pontryagin maximum principle to derive the best response of $i$,
\begin{align*}
    \alpha_i^*(t)=  \text{argmax} \ \{ H_i(\alpha_i(t), \alpha_{-i}(t), x_i(t), \lambda_i(t))\}.
\end{align*}
The state space is compact and convex.
Also, $H_i$ is concavity and differentiable with respect to control $\alpha_i(t)$.
We then obtain the best response Eqn.~\ref{eq: a*} by solving $\frac{\partial H_i}{\partial \alpha_i} = 0$.
\end{IEEEproof}

\iftp
\else
At each time instance, we are able to solve for explicit value of $\alpha^*_i(t)$ through numerical approach from following Pontryagin necessary conditions
\begin{align}
	\frac{\partial H_i}{\partial \alpha_i} = 0, \quad -\frac{\partial H_i}{\partial x_i} = \dot{\lambda_i}. \notag
\end{align}
\fi

\iftp
\else
We can now acquire the open-loop NE in steady status from Pontryagin necessary conditions (see our online technical report ~\cite{report} for the proof).
\fi
\begin{theorem}\label{the: a*}
	The best response of malicious node $i$ at the open-loop equilibrium is given by
	\begin{align}\label{eq: da*}
            \alpha_i^*(t) = \frac{p_i + 2r_i(1 + \sum_{j \in -i}\alpha_j(t))}
			{2[p_i+q_i+r_i + (q_i + r_i)\sum_{j \in -i}\alpha_j(t)]}.
	\end{align}
\end{theorem}

\vspace{1mm}
\noindent \textbf{Remark 1}:
In the steady status, the optimal dynamic control coincides with the static solution for the single malicious node setting ($n=1$).

\vspace{1mm}
\noindent \textbf{Remark 2}:
For $n$ non-cooperative malicious nodes, we are able to get $n$ best responses in steady status separately.
Thus, we have $n$ simultaneous equations with $n$ unknown variables.
It is trivial to obtain explicit solutions in some cases using analytical or numerical techniques.

To bound our analysis in a controllable scope, we take the situation of two symmetric players as a simple illustration.
The symmetric means that the payoff and the cost factors are same for two players.
We simply denote them as $p$, $q$ and $r$ respectably.
\begin{corollary}\label{cor: 2a*}
Consider two symmetric competing malicious nodes exist in the SNS, the optimal system maneuver is given by
\begin{align}
r^*(t) = (p + q)(3 - 2\beta(t))^2 - \frac{1}{4} (3p + q),
\end{align}
where $\beta(t)$ is the control of malicious behavior from two symmetric malicious nodes in steady status.
\end{corollary}
\begin{IEEEproof}
From Theorem (\ref{the: a*}), we know that
\begin{align*}
	\beta^*(t) = \beta^*_i(t) = \beta^*_{j}(t) = \frac{1}{2} (3 - \frac{\sqrt{(p + q)(3p + q + 4r)}}{2(p + q)}).
\end{align*}
which can be used to derive the system maneuver $r^*(t)$.
\end{IEEEproof}
\vspace{1mm}
We can further obtain the following corollary for general situations.
\begin{corollary}
Consider the game among two non-cooperative players $i$ and $j$.
At the open-loop equilibrium, the best response function of $i$ is negatively sloped for all $p_{i}, q_{i}$ and $r_{i} \in (0, \infty]$. In absolute value, the slope is everywhere decreasing in $r_{i}$.
\end{corollary}
\begin{IEEEproof}
The slope of best response function at the open-loop equilibrium is given by
\begin{align*}
\frac{\partial\alpha_{i}}{\partial\alpha_{j}} = -\frac{4 r_i (q_i + r_i) + p_i (q_i + 3 r_i)}{2 (p_i + (q_i + r_i) (1 + \alpha_{j})^2)} < 0.
\end{align*}
\end{IEEEproof}

%% file: simulation.tex
\section{Numerical Study}\label{sec:simulation}
In this section, we illustrate the results with numerical examples.
We build our simulation on Matlab platform with bvp4c toolbox.

\begin{figure*}[!htbp]
  \captionsetup{justification=centering}
  \begin{center}
    \subfigure[state trajectory for two identical players]{
    \label{fig: gduop}
    \includegraphics[width=0.3\linewidth]{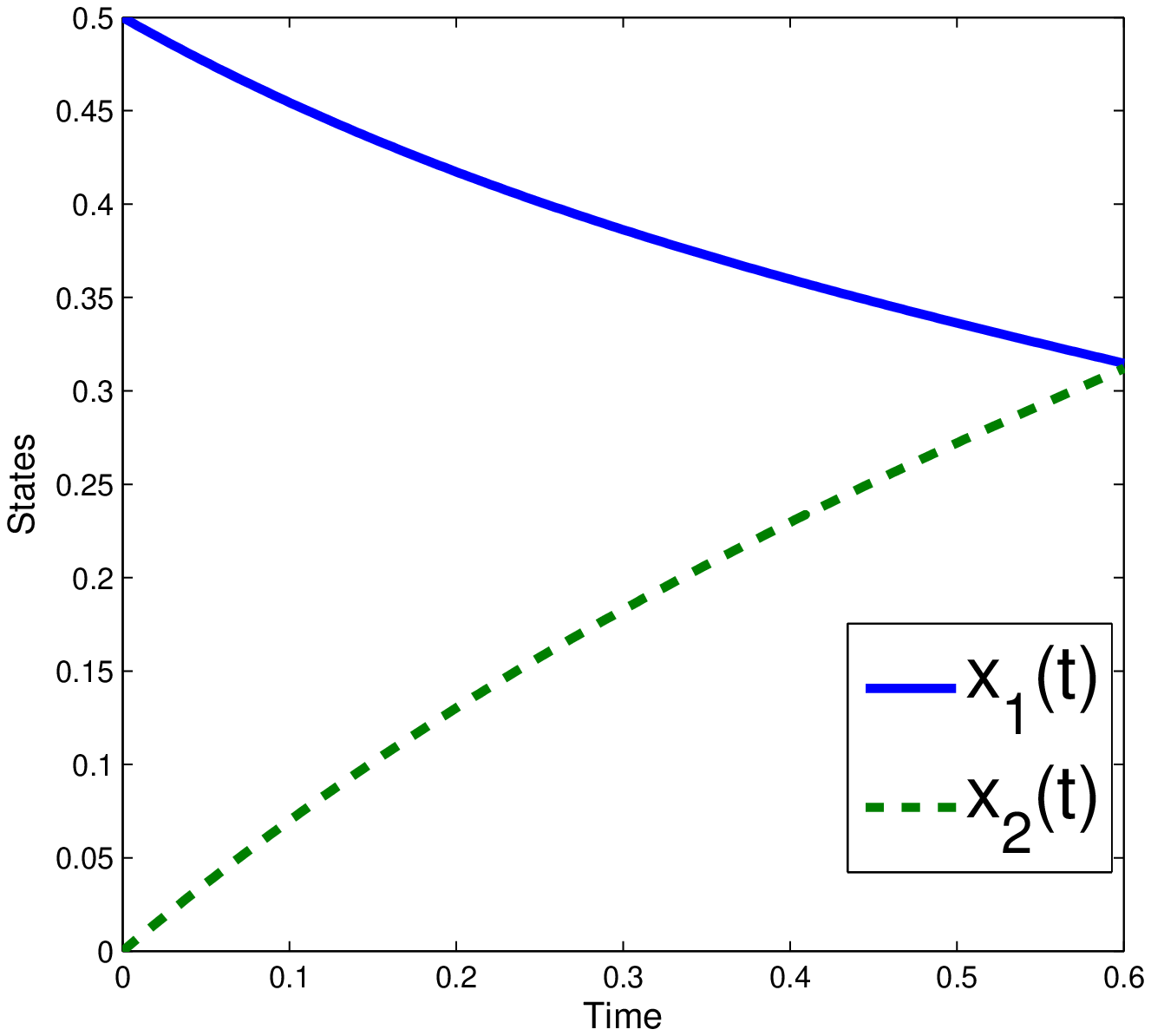}}
    \subfigure[variation of $\alpha^*$ and $\beta^*$ when $n$ increases]{
    \label{fig: comp}
    \includegraphics[width=0.3\linewidth]{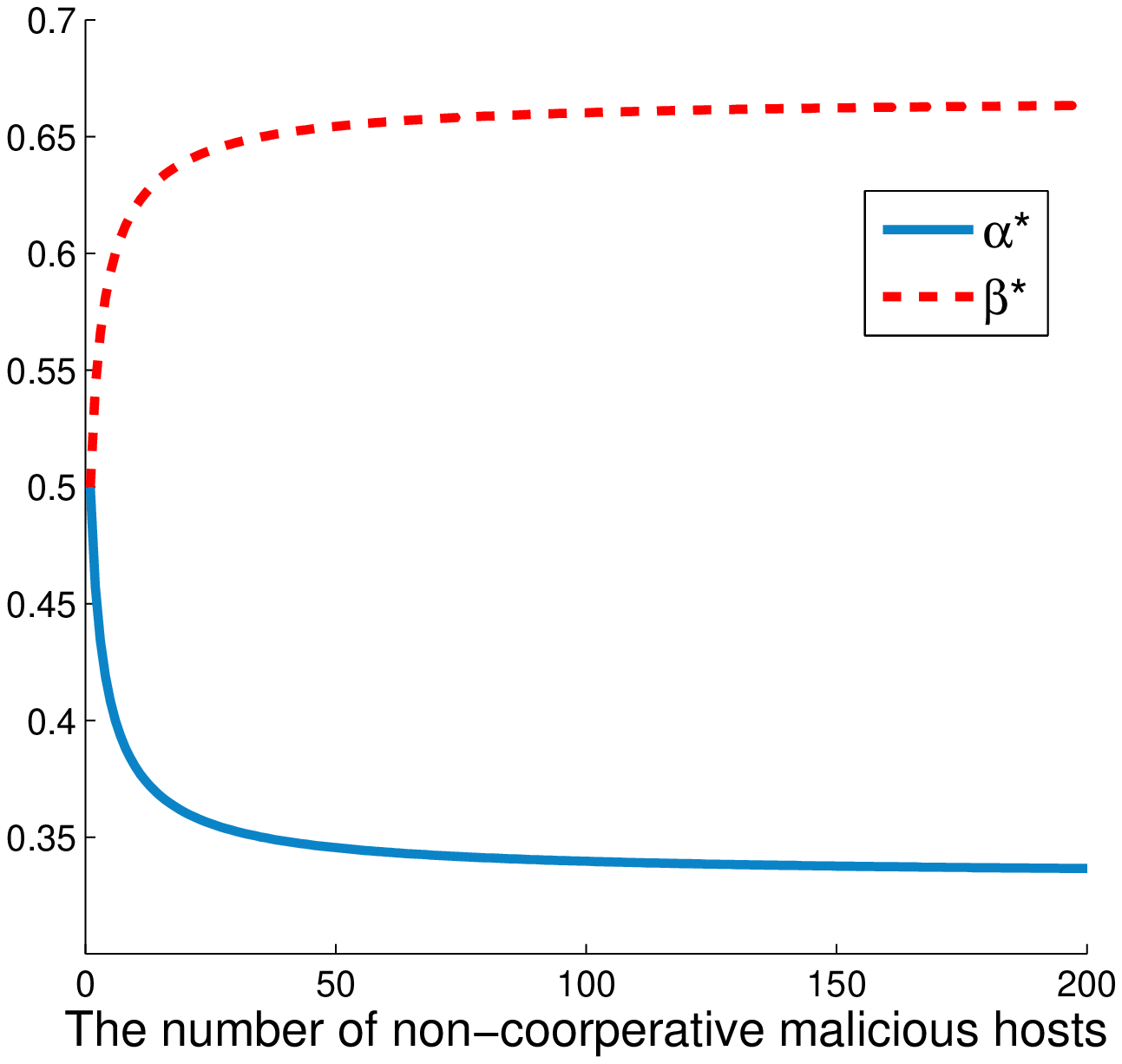}}
    \subfigure[trajectory of the states and the controls for two players with different system maneuvers]{
    \label{fig: gd2u}
    \includegraphics[width=0.3\linewidth]{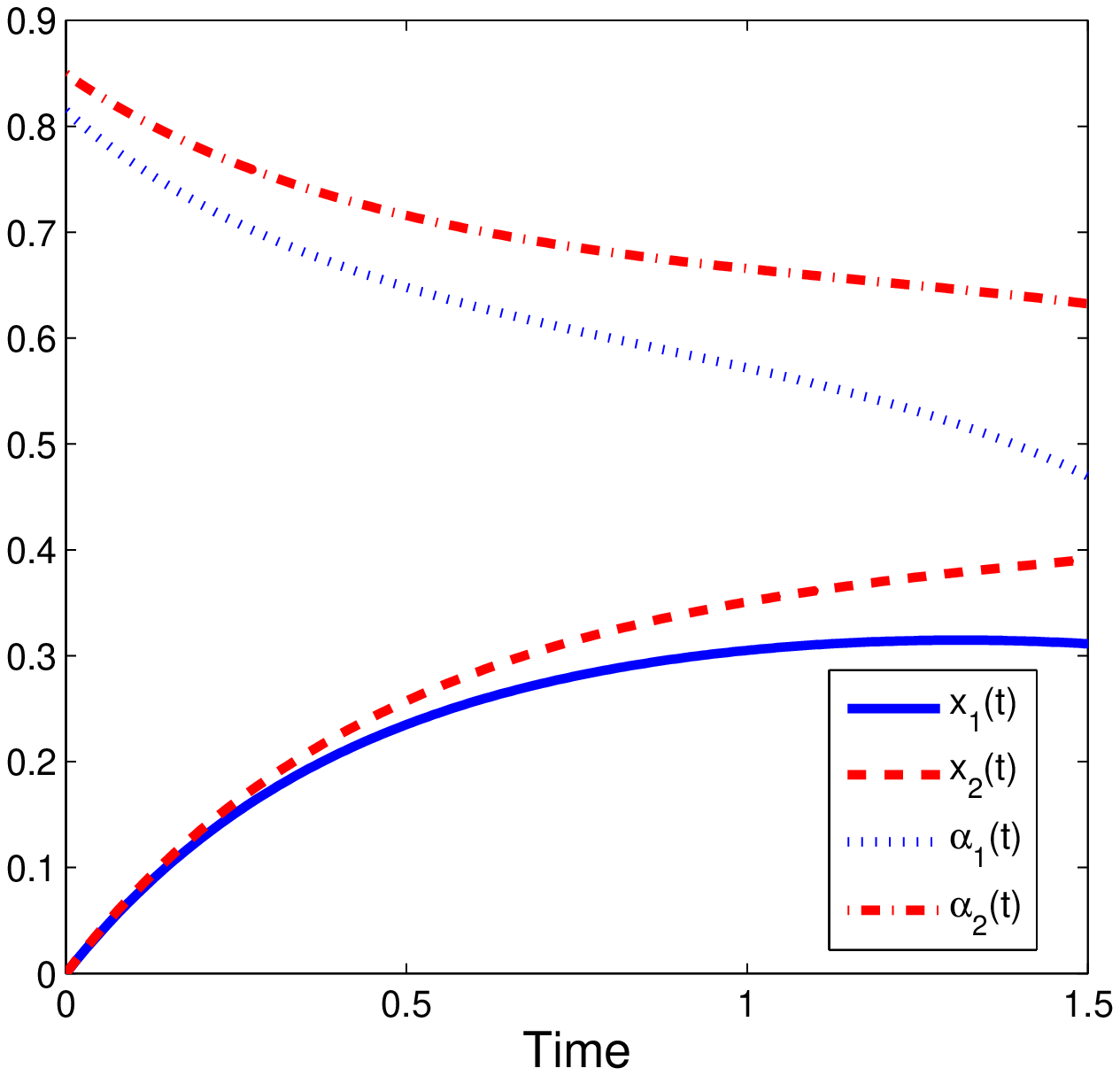}}
    \caption{Multiple Competing Malicious Nodes}
    \label{fig: q2r3}
  \end{center}
\end{figure*}

Suppose there is an existing malicious node that has already reached its steady state.
Now we introduce another homogenous malicious node with identical configurations with the existing node.
Let $p = 0.4, q = 0.2, r = 0.2$ for both nodes.

From Fig.~\ref{fig: gduop}, we can observe that the player I deviates from its previously steady state $0.5$ and its $x_1(t)$ begins decreasing, meanwhile, $x_2(t)$ of player II starts from 0 and increases until finally converging to the steady position, which matches the analytical result for steady position.
Since the factors of two players are symmetric, it is not surprise that two players finally converge to the same state.

Next, we examine how the control in the steady state evolves when the amount of players increases.
The parameters are set as same, \emph{i.e.}, $p = 0.4$ and $q = r = 0.2$.
As shown in Fig.~\ref{fig: comp}, $\alpha^*(t)$ begins at $0.5$ and converges to $0.35$ when $n$ increases, whereas $\beta^*$ starts from $0.5$ and converges to $0.65$. This observation means that the competition does not motivate good behaviors by nodes.

We then study the influences of the system maneuver $r_i$ on the controls and the states of a two-player game scenario.
Let $p_1 = p_2 = 0.5$ and $q_1 = q_2 = 0.1$, Fig.~\ref{fig: gd2u} depicts the evolution progress of the states and the controls of two players with $r_1 = 0.2$ and $r_2 = 0.3$ separately.
We can see that the higher system maneuver comes with the lower negative activity rate in social trust games.

%% file: conclusion.tex
\section{Conclusion}
\label{sec:conclusion}
This paper investigates the social-trust-based information dissemination by malicious nodes in social network sites. An interaction-based social trust model is presented. For studying the best response strategies of malicious nodes with a long-term objective, we formulate the maximization of malicious gains in a long time-span of multiple competing malicious nodes as a non-cooperative differential game. Through rigorous analysis, optimal response strategies for each malicious node are identified and the best system maneuver mechanisms are presented in order to achieve the targeted level of overall malicious activities in the system. The numerical studies further verify the analytical results.

%% file: main.bbl
\begin{thebibliography}{10}

\bibitem{ellison2007social}
N.~B. Ellison {\em et~al.}, ``Social network sites: Definition, history, and
  scholarship,'' {\em Journal of Computer-Mediated Communication}, vol.~13,
  no.~1, pp.~210--230, 2007.

\bibitem{boshmaf2011socialbot}
Y.~Boshmaf, I.~Muslukhov, K.~Beznosov, and M.~Ripeanu, ``The socialbot network:
  when bots socialize for fame and money,'' in {\em ACSAC}, 2011.

\bibitem{thomas2011design}
K.~Thomas, C.~Grier, J.~Ma, V.~Paxson, and D.~Song, ``Design and evaluation of
  a real-time url spam filtering service,'' in {\em SP}, 2011.

\bibitem{sherchan2013survey}
W.~Sherchan, S.~Nepal, and C.~Paris, ``A survey of trust in social networks,''
  {\em ACM Computing Surveys (CSUR)}, vol.~45, no.~4, 2013.

\bibitem{stringhini2010detecting}
G.~Stringhini, C.~Kruegel, and G.~Vigna, ``Detecting spammers on social
  networks,'' in {\em ACSAC}, 2010.

\bibitem{yang2011free}
C.~Yang, R.~C. Harkreader, and G.~Gu, ``Die free or live hard? empirical
  evaluation and new design for fighting evolving twitter spammers,'' in {\em
  Recent Advances in Intrusion Detection}, pp.~318--337, Springer, 2011.

\bibitem{wang2014provenance}
X.~Wang, H.~Fu, C.~Xu, and P.~Mohapatra, ``Provenance logic: Enabling
  multi-event based trust in mobile sensing,'' in {\em IPCCC}, 2014.

\bibitem{hu2015dynamic}
P.~Hu, H.~Li, H.~Fu, D.~Cansever, and P.~Mohapatra, ``Dynamic defense strategy
  against advanced persistent threat with insiders,'' in {\em INFOCOM}, 2015.

\bibitem{xiaotao2016}
X.~Feng, Z.~Zheng, D.~Cansever, A.~Swami, and P.~Mohapatra, ``Stealthy attacks
  with insider information: A game theoretic model with asymmetric feedback,''
  in {\em MILCOM}, 2016.

\bibitem{xiaotaomilcom}
X.~Feng, Z.~Zheng, P.~Hu, D.~Cansever, and P.~Mohapatra, ``Stealthy attacks
  meets insider threats: A three-player game model,'' in {\em MILCOM}, 2015.

\bibitem{omic2009protecting}
J.~Omic, A.~Orda, and P.~Van~Mieghem, ``Protecting against network infections:
  A game theoretic perspective,'' in {\em INFOCOM}, 2009.

\bibitem{6426481}
Q.~Zhu, L.~Bushnell, and T.~Basar, ``Game-theoretic analysis of node capture
  and cloning attack with multiple attackers in wireless sensor networks,'' in
  {\em CDC}, 2012.

\bibitem{jiang2013optimally}
B.~Jiang, N.~Hegde, L.~Massouli{\'e}, and D.~Towsley, ``How to optimally
  allocate your budget of attention in social networks,'' in {\em INFOCOM},
  2013.

\bibitem{jorgensen1982survey}
S.~J{\o}rgensen, ``A survey of some differential games in advertising,'' {\em
  Journal of Economic Dynamics and Control}, vol.~4, pp.~341--369, 1982.

\bibitem{han2012game}
Z.~Han, {\em Game theory in wireless and communication networks: theory,
  models, and applications}.
\newblock Cambridge University Press, 2012.

\bibitem{report}
H.~Fu, H.~Li, Z.~Zheng, P.~Hu, and P.~Mohapatra, ``Trust exploitation and
  attention competition: A game theoretic model.'' Technical Report, available
  online at http://spirit.cs.ucdavis.edu/pubs/tr/fu-trust-report.pdf.

\bibitem{border1990fixed}
K.~C. Border, ``Fixed point theorems with applications to economics and game
  theory,'' {\em Cambridge Books}, 1990.

\bibitem{cellini2007differential}
R.~Cellini and L.~Lambertini, ``A differential oligopoly game with
  differentiated goods and sticky prices,'' {\em European Journal of
  Operational Research}, vol.~176, no.~2, pp.~1131--1144, 2007.

\end{thebibliography}
